\documentclass{article}
\usepackage{amsmath}

\input{tcilatex}
\begin{document}

\title{Second Order Corrections to the Magnetic Moment of Electron at Finite
Temperature}
\author{{\small \ Samina S. Masood}$^{\ast }$ {\small and Mahnaz Q. Haseeb}$%
^{\ast \ast }${\small \ } \\
$^{\ast }${\small Department of Physics, University of Houston Clear Lake,
Houston TX 77058, }\\
{\small masood@uhcl.edu; }\\
$^{\ast \ast }${\small Department of Physics, COMSATS Institute of
Information Technology, Islamabad }\\
{\small mahnazhaseeb@comsats.edu.pk.}}
\maketitle

\begin{abstract}
Magnetic moment of electron at finite temperature is directly related to the
modified electron mass in the background heat bath. Magnetic moment of
electron gets modified at finite temperature also, when it couples with the
magnetic field, through its temperature dependent physical mass. We show
that the second order corrections to the magnetic moment of electron is a
complicated function of temperature. We calculate the self-mass induced
thermal contributions to the magnetic moment of electron, up to the two loop
level, for temperatures valid around the era of primordial nucleosynthesis.
A comparison of thermal behavior of the magnetic moment is also
quantitatively studied in detail, around the temperatures below and above
the nucleosynthesis temperature.
\end{abstract}

\section{Introduction}

Quantum Electrodynamics (QED) is well known as the simplest representative
and the most accurate gauge theory. Thermal medium effects are incorporated
in QED by taking into account the vacuum fluctuations of propagating
particles along with the hot particles in the background heat bath. For
simplicity, all the particles in the background are assumed to be in thermal
equilibrium with the heat bath. These particles are virtually created and
annihilated continuously due to the effects of heat bath at finite
temperature. The interactions with the background electrons, positrons and
photons are included through statistical distribution functions of fermions
and bosons, known as Fermi-Dirac distribution and Bose-Einstein
distribution. These distribution functions represent the possibility of
exchange of virtual particles with the real hot particles from the heat
bath. Electromagnetic interactions of particles get modified at finite
temperature because of the many-body aspects of the statistical background
possessed by the hot medium. This replaces the notion of one-particle
systems adopted for particle interactions in the vacuum with many particle
aspects.

The techniques for handling the particle interactions in the background
medium have extensively evolved over the last few decades and are a part of
standard literature (see for example [1], [2] and references therein). It
has been explicitly demonstrated [3-10] that in the hot background, the
density-of-states factors have to be modified to include the real emission
and absorption of particles which are in thermal equilibrium with the heat
bath. The validity of QED renormalization in a background of particles is
refined by comparing the second order (in $\alpha $) corrections with the
first order corrections from the heat bath. It is explicitly seen, as
expected, that one-loop radiative corrections are significantly larger than
the two-loop corrections. Our scheme of calculations is based on the real
part of the propagator and the results are valid, at least below the
decoupling temperature [11], i.e., around 2 MeV.

Renormalization techniques of vacuum theory are extended to include finite
temperature effects in a standard manner. Regular renormalization procedure
for QED in vacuum is used at finite temperature to study the background
effects on electron mass, charge and wave function renormalization
constants. The modifications in the electromagnetic properties are estimated
in terms of the renormalized values of QED parameters up to the two loop
level [3-19].

Feynman rules at finite temperature remain the same as those in vacuum
except that the particle propagators are appropriately modified. We work in
Minkowski space where the Green's functions depend on real Minkowski momenta 
$p^{\mu }$. Therefore, the dynamical processes such as particles propagating
in the heat bath may be more conveniently dealt with. Moreover, in the
real-time formulation, the thermal corrections can be easily kept separate
from the vacuum corrections. Therefore, order by order cancellation of
temperature dependent singularities, and the convergence of perturbative
expansion can be straight away tracked down.

In this paper, thermal contributions to the anomalous magnetic moment of
electron are specifically studied. The magnetic moment of electron modifies
through the radiative corrections to the electron mass both at the one loop
and two loop levels. We analyze net effect from the first order and the
second order radiative contributions, including both the irreducible and
disconnected graphs up to the two loop level. Second order selfmass
corrections due to finite temperature (of the order $\alpha ^{2}$) are used
here to estimate finite temperature effects on the magnetic moment of
electron.

\section{Self-Mass of an Electron at Finite Temperature}

The renormalized mass of electron is represented by a physical mass given as

\begin{equation}
m_{phys}=m+\delta m,
\end{equation}%
where $m$ is the electron mass at zero temperature. Radiatively corrected
physical mass up to order $\alpha ^{2}$\ is

\begin{equation*}
m_{phys}\cong m+\delta m^{(1)}+\delta m^{(2)},
\end{equation*}%
with $\delta m^{(1)}$ and $\delta m^{(2)}$\ as the shifts in the electron
mass at one and two loop level respectively. The physical mass is deduced by
locating pole of the propagator $\frac{i(\NEG{p}+m)}{p^{2}-m^{2}+i%
\varepsilon }$. For this purpose, all finite terms in electron self-energy
are combined together. The physical mass of the electron at one loop was
obtained by writing

\begin{equation}
\Sigma (p)=A(p)E\gamma _{0}-B(p)\vec{p}.\vec{\gamma}-C(p),
\end{equation}%
where $A(p)$, $B(p)$, and $C(p)$ are the relevant coefficients. Taking the
inverse of the propagator with momentum and mass term separated as

\begin{equation}
S^{-1}(p)=(1-A)E\gamma _{0}-(1-B)\vec{p}.\vec{\gamma}-(m-C).
\end{equation}%
The temperature-dependent radiative corrections to the electron mass up to
the first order in $\alpha $, are obtained from the temperature modified
propagator. These corrections are rewritten in the form of boson and fermion
loop integrals at the one loop level as%
\begin{eqnarray}
E^{2}-|\mathbf{p}|^{2} &=&m^{2}+\frac{\alpha }{2\pi ^{2}}\left(
I.p+J_{B}.p+m^{2}J_{A}\right)   \notag \\
&\equiv &m_{phys}^{2},
\end{eqnarray}%
where \qquad \qquad \qquad \qquad \qquad \qquad \qquad \qquad \qquad \qquad
\qquad \qquad \qquad \qquad \qquad \qquad \qquad \qquad \qquad \qquad \qquad 
\begin{equation}
I.p=\frac{4\pi ^{3}T^{2}}{3},
\end{equation}%
and \qquad \qquad \qquad \qquad \qquad \qquad \qquad \qquad \qquad \qquad
\qquad \qquad \qquad \qquad \qquad \qquad \qquad \qquad 
\begin{equation}
J_{B}.p=8\pi \left[ \frac{m}{\beta }a(m\beta )-\frac{m^{2}}{2}b(m\beta )-%
\frac{1}{\beta ^{2}}c(m\beta )\right] .
\end{equation}%
Thus up to the first order in $\alpha ,$\ thermal corrections to the mass of
electron were obtained in ref. [3] to be 
\begin{equation}
m_{phys}^{2}=m^{2}\left[ 1-\frac{6\alpha }{\pi }b(m\beta )\right] +\frac{%
4\alpha }{\pi }mT\text{ }a(m\beta )+\frac{2}{3}\alpha \pi T^{2}\left[ 1-%
\frac{6}{\pi ^{2}}c(m\beta )\right] .
\end{equation}%
The first order correction at finite temperature is calculated as 
\begin{eqnarray}
\frac{\delta m}{m} &\simeq &\frac{1}{2m^{2}}\left( m_{phys}^{2}-m^{2}\right) 
\notag \\
&\simeq &\frac{\alpha \pi T^{2}}{3m^{2}}\left[ 1-\frac{6}{\pi ^{2}}c(m\beta )%
\right] +\frac{2\alpha }{\pi }\frac{T}{m}a(m\beta )-\frac{3\alpha }{\pi }%
b(m\beta ),
\end{eqnarray}%
with $\frac{\delta m}{m}$ as the relative shift in electron mass due to
finite temperature which was originally determined in ref. [3] with%
\begin{equation}
a(m\beta )=\ln (1+e^{-m\beta }),
\end{equation}%
\begin{equation}
b(m\beta )=\dsum\limits_{n=1}^{\infty }(-1)^{n}\func{Ei}(-nm\beta ),
\end{equation}%
\begin{equation}
c(m\beta )=\dsum\limits_{n=1}^{\infty }(-1)^{n}\frac{e^{-nm\beta }}{n^{2}},
\end{equation}%
At low temperature, the functions $a(m\beta )$, $b(m\beta )$, and $c(m\beta )
${\LARGE \ }fall off in powers of $e^{-m\beta }$ in comparison with $\left( 
\frac{T}{m}\right) ^{2}$ and can be neglected so that%
\begin{equation}
\frac{\delta m}{m}\overset{T\ll m}{\longrightarrow }\frac{\alpha \pi T^{2}}{%
3m^{2}}.
\end{equation}%
Moreover, in the high-temperature limit, $a(m\beta )$ and $b(m\beta )$ are
vanishingly small whereas $c(m\beta )\longrightarrow -\pi ^{2}/12$, yielding%
\begin{equation}
\frac{\delta m}{m}\overset{T>m}{\longrightarrow }\frac{\alpha \pi T^{2}}{%
2m^{2}}.
\end{equation}%
Eq. (8) is valid for large temperatures relevant in QED including $T\sim $ $%
m.$ This range of temperature is particularly interesting\ from the point of
view of primordial nucleosynthesis. It has been found that some parameters
in the early universe such as the energy density and the helium abundance
parameter $Y$ become slowly varying functions of temperature [20-22] whereas
they remain constant in both extreme limits given by $T\ll m$\ and $T\gg m$.

Using the same procedure as the one used for one loop calculations, the
relative shift in electron mass at the two loop level was obtained in ref.
[17]. This relative shift in electron mass introduces temperature dependence
in the magnetic moment of electron up to two loops. However, the two-loop
order result is very complicated and cannot be easily simplified. Therefore,
we will use the complete expression for the two loop calculations of
electron selfmass, near the nucleosynthesis temperature, given in refs.
[12-14]. In the following section, we compute the magnetic moment of
electron from the self-mass of electron, up to the two loop level, in
thermal background.

\section{Magnetic Moment of Electron in the Heat Bath}

The anomalous magnetic moment of an electron is generated due to the
coupling of electron with the magnetic field through the radiative
corrections. Some of these results that are used here were given in ref.
[17]. The electromagnetic coupling is affected by the electron mass and the
radiative corrections to the electron mass. The coupling of electron mass
with the external magnetic field is regulated through mass of the particle
itself. It is known from the calculation of the radiative corrections that
the self-mass corrections to the electron are contributed by the
distribution of hot bosons and fermions in the background medium. This
effect, in turn, changes the electromagnetic properties of the medium
itself. Therefore the magnetic moment gets changed with the finite
temperature effects. The magnetic moment of electron is related to the
relative shift in electron mass at finite temperature $\frac{\delta m}{m}$
as:

\begin{equation}
\mu _{a}=\frac{\alpha }{2\pi }-\frac{2}{3}\frac{\delta m}{m}.
\end{equation}

The leading order contributions to the magnetic moment up to the one loop
level is

\begin{equation}
\mu _{a}=\frac{\alpha }{2\pi }-\ \frac{2}{3}\alpha \left[ \frac{\pi T^{2}}{%
3m^{2}}\left \{ 1-\frac{6}{\pi }c(m\beta )\right \} +\frac{2}{\pi }\frac{T}{m%
}a(m\beta )-b(m\beta )\right]
\end{equation}%
which can be shown to be: 
\begin{equation}
\mu _{a}=\frac{\alpha }{2\pi }-\ \frac{2}{9}\frac{\alpha \pi T^{2}}{m^{2}}
\end{equation}%
for $T<m$ while it becomes:

\begin{equation}
\mu _{a}=\frac{\alpha }{2\pi }-\ \frac{1}{3}\frac{\alpha \pi T^{2}}{m^{2}}
\end{equation}%
for $T>m.$ First order in $\alpha $ contribution to $\mu _{a}$ around the
temperature range relevant for primordial nucleosynthesis (i.e., $T\sim m$),
soon after the big bang is given by expression in eq. (15). The two loop
contribution to the magnetic moment can simply be added to the magnetic
moment in terms of the relative shift in electron mass $\frac{\delta m^{(2)}%
}{m}$ as

\begin{equation}
\mu _{a}=\frac{\alpha }{2\pi }-\frac{2}{3}\left( \frac{\delta m^{(1)}}{m}+%
\frac{\delta m^{(2)}}{m}\right) .
\end{equation}

Now using the expression for $\frac{\delta m}{m}$ at finite temperature in
ref. [14], we get thermal contributions to the magnetic moment up to the
two-loop level as

\begin{eqnarray}
\mu _{a} &=&\frac{\alpha }{2\pi }-\ \frac{2}{3}\alpha \left[ \frac{\pi T^{2}%
}{3m^{2}}\left \{ 1-\frac{6}{\pi }c(m\beta )\right \} +\frac{2}{\pi }\frac{T%
}{m}a(m\beta )-\frac{3}{\pi }b(m\beta )\right]  \notag \\
&&-\frac{2\alpha ^{2}}{3}\left[ \frac{\pi T^{2}}{3m^{2}}\left \{ 1-\frac{6}{%
\pi }c(m\beta )\right \} +\frac{2}{\pi }\frac{T}{m}a(m\beta )-\frac{3}{\pi }%
b(m\beta )\right]  \notag \\
&&\times \left[ \frac{\pi T^{2}}{3m^{2}}\left \{ 1-\frac{6}{\pi }c(m\beta
)\right \} +\frac{2}{\pi }\frac{T}{m}a(m\beta )-\frac{3}{\pi }b(m\beta )%
\right]  \notag \\
&&-\frac{4}{3}\alpha ^{2}\dsum \limits_{r=1}^{\infty }[-\frac{m^{2}\beta ^{2}%
}{\pi ^{2}}c(m\beta )+T^{2}\{ \dsum \limits_{n=3}^{r+1}\text{ }(-1)^{n+r+1}%
\frac{\pi \text{ }}{6mEv}\frac{e^{-\beta (rE+mn)}}{n}  \notag \\
&&-\frac{3}{8}(-1)^{r}\frac{e^{-r\beta E}}{E^{2}v^{2}}[\frac{9E^{2}}{2m^{2}}%
+6\dsum \limits_{s=3}^{r+1}\frac{1}{s}+4\dsum \limits_{n,s=3}^{r+1}\frac{1}{%
ns}+(-1)^{s-r}\{ \frac{9E}{m}\left( 3+4\dsum \limits_{s=3}^{r+1}\frac{1}{s}%
\right)  \notag \\
&&+2\left( \frac{E^{2}v^{2}}{m^{2}}-3\right) \left( 9+18\dsum
\limits_{s=3}^{r+1}\frac{1}{s}+8\dsum \limits_{n,s=3}^{r+1}\frac{1}{ns}%
\right) \}]+\frac{4}{E^{2}v^{2}}\}  \notag \\
&&-\frac{T}{m}\{ \frac{\pi m\text{ }}{6Ev}\dsum \limits_{s=2}^{r+1}\dsum
\limits_{n=1}^{s+1}\frac{e^{-\beta (rE+mn)}}{n}\left[ 1\ -\left \{
(-1)^{r+n}-(-1)^{s+n}\right \} \right]  \notag \\
&&+[\left \{ \func{Ei}(-m\beta )-\func{Ei}(-2m\beta )\right \} \{ \frac{9E}{4%
}\left( \frac{E}{E^{2}v^{2}}-\frac{1}{m}\right)  \notag \\
&&+\left( \frac{5E}{m}-21+\frac{E^{2}}{2m^{2}}\right) \dsum
\limits_{n=3}^{r+1}\frac{1}{n}\}  \notag \\
&&+\left \{ \frac{9}{4v^{2}}-\dsum \limits_{n=1}^{s+1}\left[ \dsum
\limits_{s=3}^{r+1}1-E^{2}\left( \frac{1}{2m^{2}}+\frac{3}{E^{2}v^{2}}%
\right) +\frac{3E}{m}\right] \right \} (-1)^{s}\func{Ei}\text{ }(-sm\beta )]
\notag \\
&&+\frac{e^{-rm\beta }}{m}\{ \left[ \frac{9E}{2v^{2}}+2\left( \frac{3E}{v^{2}%
}+\frac{3E^{2}v^{2}}{m}-5E\right) \dsum \limits_{n=3}^{r+1}\frac{1}{n}\right]
\dsum \limits_{s=1}^{\infty }\sinh sm\beta  \notag \\
&&-\frac{3m^{3}}{E^{2}v^{2}}\left( \frac{3}{4}-\dsum \limits_{n=3}^{r+1}%
\frac{1}{n}\right) \dsum \limits_{s=1}^{\infty }\cosh sm\beta \}]\}+\frac{1}{%
m^{2}}\{ \frac{9m}{4E^{2}v^{2}}\left( E^{3}+\frac{m^{3}}{2}\right)  \notag \\
&&+\left[ \frac{3m}{E^{2}v^{2}}(E^{3}+m^{3})+5mE-3E^{2}v^{2}\right] \dsum
\limits_{n=3}^{r+1}\frac{1}{n}\} \left \{ \func{Ei}(-m\beta )-2\func{Ei}%
(-2m\beta )\right \}  \notag \\
&&-\frac{1}{m^{2}}\dsum \limits_{n=3}^{r+1}\{ \dsum \limits_{s=1}^{r+1}\frac{%
(-1)^{s}}{n}[\frac{m^{2}re^{-sm\beta }}{2}  \notag \\
&&+\left \{ sE\left( 2m-\frac{E^{2}}{m}\right) +\frac{m^{2}(s-r)}{2}\right
\} \func{Ei}(-sm\beta )]  \notag \\
&&-\frac{\pi m^{3}\text{ }}{3Ev}\left[ e^{-\beta rE}\text{ }(-1)^{n+r}(n+1)\
-\dsum \limits_{s=2}^{r+1}(-1)^{n+s}\right] \func{Ei}(-nm\beta )\}].
\end{eqnarray}

It can be clearly seen from eq. (19)\ that the second order corrections are
suppressed by at least two orders of magnitudes as compared to the one loop
contributions. Dependence of the\textbf{\ }self-mass induced thermal
contributions to the anomalous magnetic moment of electron is very
complicated at the two loop level, as indicated by eq. (19). The exact
estimate of this magnetic moment for application to the primordial
nucleosynthesis is very involved and probably is not really so significant
at two loop level. However, low temperature $(T<m)$ and high temperature $%
(T>m)$\ values of the magnetic moment can be quantitatively analyzed to
prove the validity of the renormalization scheme. We give the plotting for
magnetic moment $\mu _{a}$ vs $\frac{T}{m}$ in low temperature and the high
temperature regions in the next section. This analysis indicates that the
magnetic moment of electron changes its behavior around nucleosynthesis.

\section{Results and Discussion \ \ }

The electron mass acquires a significant contribution from the heat bath
even for temperatures that are smaller than the electron mass. However, this
dependence becomes very complicated as soon as the background temperature
approaches the value of electron mass. One loop corrections to the electron
mass at finite temperature are presented in eq. (8). The low ($T<m$) and
high ($T>m$) temperature values of self-mass of electron at the one loop
level are given in eqs. (12) and (13), respectively as limiting cases of eq.
(8). Incorporating the second order relation for the physical mass of
electron [17] into eq. (14) leads to eq. (19) which gives the general form
of the anomalous magnetic moment at finite temperature up to order $\alpha
^{2}$. When an electron couples with the magnetic field at finite
temperature, a nonzero contribution to the magnetic moment is picked up due
to the coupling of electron mass with the thermal background. Eq. (14)
presents the acceptable relation of the magnetic moment with that of the
self-mass of electron.

\FRAME{ftbpFU}{5.0825in}{3.3157in}{0pt}{\Qcb{Low temperature behavior of the
magnetic moment of electron at the two loop level.}}{}{%
t_less_than_m_mag_mom.eps}{\special{language "Scientific Word";type
"GRAPHIC";maintain-aspect-ratio TRUE;display "USEDEF";valid_file "F";width
5.0825in;height 3.3157in;depth 0pt;original-width 5.028in;original-height
3.2707in;cropleft "0";croptop "1";cropright "1";cropbottom "0";filename
'T_Less_than_m_mag_mom.eps';file-properties "XNPEU";}}

\FRAME{ftbpFU}{5.0955in}{3.064in}{0pt}{\Qcb{High temperature behavior of the
magnetic moment of electron at the two loop level.}}{}{tgtm_mag_mom.eps}{%
\special{language "Scientific Word";type "GRAPHIC";maintain-aspect-ratio
TRUE;display "USEDEF";valid_file "F";width 5.0955in;height 3.064in;depth
0pt;original-width 5.0401in;original-height 3.0199in;cropleft "0";croptop
"1";cropright "1";cropbottom "0";filename 'TGtM_mag_mom.eps';file-properties
"XNPEU";}}

The behavior of the magnetic moment of electron near the nucleosynthesis
temperatures (eq.(19)) is very complicated\ and it can be fitted through a
single mathematical function under some special conditions only. However, we
can extract the quantitative behavior of magnetic moment of electron for low
and high temperatures in a comparatively simple form. Using previously
studied second order contributions to the electron mass at low temperature $%
(T<m)$, leading order contributions to the magnetic moment of electron can
be computed as

\begin{equation}
\mu _{a}\overset{T<m}{\longrightarrow }-\ \frac{2}{9}\frac{\alpha \pi T^{2}}{%
m^{2}}-10\alpha ^{2}\left( \frac{T^{2}}{m^{2}}\right) ,
\end{equation}%
whereas, the leading order contributions at high temperature $(T>m)$ comes
out to be

\begin{equation}
\mu _{a}\overset{T>m}{\longrightarrow }-\ \frac{1}{3}\frac{\alpha \pi T^{2}}{%
m^{2}}-\frac{\alpha ^{2}\pi ^{2}}{6}\left( \frac{T^{2}}{m^{2}}\right) ^{2}+%
\frac{\alpha ^{2}}{6}\frac{m^{2}}{T^{2}}.
\end{equation}%
Eqs. (20) and (21) are used for a quantitative study of magnetic moment of
electron at low temperature and high temperature, respectively. We plot the
temperature dependence of magnetic moment of electron versus $\frac{T}{m}.$\
A plot of eq. (20) is given in fig. 1, whereas, eq. (21) is plotted in fig.
2. Both of these graphs give a sort of quadratic behavior in the negative
sense.

Eqs. (21) and (22), fig. 1 and fig. 2. indicate the difference between low
temperature and high temperature behavior. Major difference in the behavior
occurs due to the $m^{2}/T^{2}$ term at high temperature. Contribution of
this term reduces with increasing temperatures and becomes totally ignorable
at very high temperatures. This is obvious from fig. 1 and fig. 2 that the
magnetic moment of electron falls off rapidly with temperature after
nucelosynthesis as compared to that before nucleosynthesis. This rapid
decrease in magnetic moment after the nucleosynthesis is not the same as it
is compared in eqs. (16) and (17), implying that the one loop and two loop
behaviors are not exactly similar.

\bigskip \textbf{References }

\begin{enumerate}
\item J.I. Kapusta, C. Gale, Finite Temperature Field Theory, (Cambridge
University Press, New York, 2006).

\item P. Landsman, Ch.G. Weert, Phys. Rep. \textbf{145}, 141 (1987).

\item K. Ahmed, Samina (Saleem) Masood, Phys. Rev. D \textbf{35,} 1861
(1987).

\item K. Ahmed, Samina (Saleem) Masood, Phys. Rev. D \textbf{35,} 4020
(1987).

\item K. Ahmed, S.S. Masood, Ann. Phys. (N.Y.) \textbf{207,} 460 (1991).

\item S.S. Masood, Phys. Rev. D \textbf{44,} 3943 (1991).

\item S.S. Masood, Phys. Rev. D \textbf{47,} 648 (1993).

\item Samina S.Masood, Mahnaz Q. Haseeb, Astropart. Phys. \textbf{3,} 405
(1995).

\item Samina S. Masood, Mahnaz Qader, Phys. Rev. D \textbf{46,} 5110 (1992).

\item Samina S. Masood, Mahnaz Qader, in \textit{Proceedings of 4th Regional
Conference on Mathematical Physics}, edited by F. Ardalan, H. Arafae, S.
Rouhani (Sharif University of Technology Press, Tehran, 1990) pp. 334.

\item Samina Masood, QED\ Near Decoupling Temperature, arXiv:1205.2937
[hep-ph] .

\item Mahnaz Qader, Samina S. Masood, K. Ahmed, Phys. Rev. D \textbf{44,}
3322 (1991).

\item Mahnaz Qader, Samina S. Masood, K. Ahmed, Phys. Rev. D \textbf{46,}
5633 (1992).

\item Mahnaz Q. Haseeb, Samina S. Masood, Chin. Phys. C \textbf{35,} 608
(2011).

\item Samina S. Masood, Mahnaz Q. Haseeb, Int. J. Mod. Phys. A \textbf{23,}
4709 (2008).

\item M.Q. Haseeb, S.S. Masood, Finite Temperature Two Loop Corrections to
Photon Self Energy, arXiv:1110.3447 [hep-th] .

\item M.Q. Haseeb, S.S. Masood, Phys. Lett. B \textbf{704,} 66 (2011).

\item M.E. Carrington, A. Gynther, P. Aurenche, Phys. Rev. D \textbf{77},
045035 (2008).

\item M.E. Carrington, A. Gynther, D. Pickering, Phys. Rev. D \textbf{78},
045018 (2008).

\item Samina (Saleem) Masood, Phys. Rev. D \textbf{36,} 2602 (1987).

\item T. Yabuki, A. Kanazawa, Prog. Theor. Phys. \textbf{85,} 381 (1991).

\item Samina Masood, QED at Finite Temperature and Density, (Lambert
Academic Publication, March, 2012).
\end{enumerate}

\end{document}